%% file: acl_latex.tex
\pdfoutput=1

\documentclass[11pt]{article}

\usepackage[final]{acl}

\usepackage{times}
\usepackage{latexsym}

\usepackage[T1]{fontenc}
\usepackage[export]{adjustbox} 

\usepackage[utf8]{inputenc}

\usepackage{microtype}

\usepackage{inconsolata}
\usepackage{float}

\usepackage{graphicx}

\usepackage{caption}
\usepackage{subcaption}
\usepackage{listings}
\usepackage{geometry}
\usepackage{booktabs}
\usepackage{multirow}
\usepackage{fvextra}
\usepackage{longtable}
\DefineVerbatimEnvironment{Verbatim}{Verbatim}{breaklines=true,breakanywhere=true}
\newcommand{\method}{Idea Novelty Checker}

\title{Literature-Grounded Novelty Assessment of Scientific Ideas}

\author{
  Simra Shahid\textsuperscript{1} \quad 
  Marissa Radensky\textsuperscript{2} \quad
  Raymond Fok\textsuperscript{2} \quad \\
  \textbf{Pao Siangliulue\textsuperscript{3} \quad
  Daniel S. Weld\textsuperscript{3} \quad
  Tom Hope\textsuperscript{3}} \\
  \textsuperscript{1}Microsoft \quad
  \textsuperscript{2}University of Washington \quad
  \textsuperscript{3}Allen Institute for AI \\
  \texttt{simrashahid@microsoft.com} \quad
  \texttt{\{radensky, rayfok\}@cs.washington.edu} \quad \\
  \texttt{\{paos, danw, tomh\}@allenai.org}
}

\begin{document}
\maketitle
\begin{abstract}

Automated scientific idea generation systems have made remarkable progress, yet the automatic evaluation of idea novelty remains a critical and underexplored challenge. Manual evaluation of novelty through literature review is labor-intensive, prone to error due to subjectivity, and impractical at scale. To address these issues, we propose the \textbf{\method}, an LLM-based retrieval-augmented generation (RAG) framework that leverages a two-stage  retrieve-then-rerank approach. The \method first collects a broad set of relevant papers using keyword and snippet-based retrieval, then refines this collection through embedding-based filtering followed by facet-based LLM re-ranking. It incorporates expert-labeled examples to guide the system in comparing papers for novelty evaluation and in generating literature-grounded reasoning. Our extensive experiments demonstrate that our novelty checker achieves approximately 13\% higher agreement than existing approaches. Ablation studies further showcases the importance of the facet-based re-ranker in identifying the most relevant literature for novelty evaluation.

\end{abstract}

\input{paper/introduction}
\input{paper/relatedwork}

\input{paper/formative}

\input{paper/approach}

\input{paper/implementation}
\input{paper/experiments}

\input{paper/ablation}

\input{paper/qualitative}

\input{paper/prompt_sensitivity}

\input{paper/conclusion}

\bibliography{references}

\appendix

\input{paper/appendix}

\end{document}

%% file: paper/introduction.tex
\section{Introduction}

Novelty evaluation is foundational for determining whether ideas in scientific research, product development, or creative ideation introduce meaningful innovation relative to prior work. Yet, as the volume of published literature grows exponentially, manual verification of originality becomes impractical. This is further complicated by the inherent subjectivity of novelty judgments, which is why experts can more easily decide on similarity of two ideas \cite{picard_concept_2023} and often struggle to articulate why one idea is more novel than another. Further the evaluation becomes subjective as it also depends on personal knowledge and intuition gained from scientific literature \cite{ahmed_interpreting_2019, picard_concept_2023}. 

Automated systems attempt to address this challenge by defining novelty as differences observed while comparing new ideas against prior work with similarity measures, but they exhibit critical limitations. Prior work has evolved from using n-gram frequency and lexical metrics (TF-IDF, LDA) \cite{wang_towards_2019, sarica_technet_2020} to semantic embeddings \cite{gomez-perez_artificial_2022, su_many_2024} that capture similarity but don't capture paraphrased variations of ideas and papers. Moreover, while recent approaches have adopted LLM-augmented pipelines  to generate numerical scores (1-10) \cite{bougie_generative_2024, wang_scipip_2024} or provide binary classifications (novel versus not novel) \cite{lu_ai_2024, li_chain_2024, si_can_2024,su_many_2024}, they do not ground the rationales in existing works and frequently fail to capture subtle variations in phrasing, resulting in the misclassification of well-documented ideas as novel \cite{beel_evaluating_2025, gupta_all_2025}. This shortcoming makes it difficult for researchers to distinguish novel ideas from incremental contributions or subtle cases of plagiarism \cite{gupta_all_2025}.

Moreover, all these approaches hinge on the successful retrieval of relevant literature for a given idea, a task that remains inherently challenging \cite{ mysore_multi-vector_2022, mysore_csfcube_2021, stevenson_beyond_2022, eger_pitfalls_2019, xu_rc-net_2014, mahajan_revisiting_2025}. Prior works \cite{si_can_2024, lu_ai_2024} extract keywords from the idea to search for papers, so important work can easily be missed if the relevant paper does not have the exact keyword. This undermines the reliability of the novelty evaluation process.

We address these gaps with \method, a retrieval-augmented LLM pipeline that assesses an idea's novelty by comparing it to a set of the most relevant papers. First, \method collects a broad set of relevant papers using keyword and snippet-based retrieval, as well as by retrieving papers similar to any seed papers provided. Next, an embedding-based similarity search filters this large collection, and a facet-based LLM re-ranker \cite{sun_is_2023} further narrows the set by comparing idea facets (purpose, mechanism, evaluation, and application) with those in the retrieved papers. Finally, expert-annotated in-context examples of \textit{novel} and \textit{not novel} ideas guide the system in generating literature-grounded rationales, mitigating subjectivity in novelty judgments.

In our experiments, we compared \method with baselines such as zero-shot prompting, prompt optimization approaches (DSPY and TextGRAD), and expert-based OpenReview examples. Our results show that expert-annotated in-context examples significantly improve classification performance. Comparisons with systems like AI Scientist and AI Researcher further demonstrate that our \method achieves higher agreement with expert judgments, and our ablation studies shows that the combined retrieval and two-stage re-ranking are critical for identifying the most relevant papers.

Our contributions are as follows:
\vspace{-2mm}
\begin{itemize}
\setlength{\itemsep}{0pt} 
    \item We introduce \method, a retrieval-augmented LLM pipeline that automatically evaluates the novelty of scientific ideas. We plan to release our code and expert-collected data\footnote{ 
    \href{https://github.com/simra-shahid/idea_novelty_checker}{github.com/simra-shahid/idea\_novelty\_checker}} to support work in automatic scientific discovery and provide literature-grounded novelty evaluations.
    
    \item We conducted a formative study in which experts evaluated ideas for novelty. The study revealed two key challenges to consider for novelty evaluation: clarifying what constitutes novelty given its subjectivity, and identifying relevant literature to assess it. This directly shaped the design of \method.

    \item Our method integrates keyword-based and snippet-based retrieval, followed by a two-stage re-ranker with embedding similarity and facet-based LLM re-ranking to identify key literature related to the given idea.

    \item We present extensive evaluations, ablation studies, and qualitative analyses that demonstrate the effectiveness of our novelty checker over existing approaches. Additionally, we discuss prompt sensitivity in LLMs for novelty evaluation further highlighting the importance of clear novelty definitions.

\end{itemize}

%% file: paper/relatedwork.tex
\section{Related Work}

Automated approaches to novelty assessment in scientific literature have evolved considerably. Early methods relied on lexical similarity metrics, such as TF-IDF, LSA, and LDA \cite{wang_towards_2019, sarica_technet_2020}, but these techniques struggled to capture paraphrased concepts. Semantic embedding methods \cite{gomez-perez_artificial_2022} improved on this by identifying deeper relationships, yet they are confined to surface-level comparisons \cite{mysore_multi-vector_2022, mysore_csfcube_2021}.

Retrieval-augmented LLM systems have emerged as a promising alternative, evaluating novelty either on a numerical scale (e.g., 1–10) \cite{bougie_generative_2024, wang_scipip_2024} or with binary classification \cite{li_chain_2024, lu_ai_2024}. AI Researcher \cite{si_can_2024} uses a Swiss-system tournament ranking to compare ideas pairwise for \textit{similarity} and \textit{novelty} against individual papers. If any comparison has sufficient similarity, the idea is not novel. Another notable work is AI Scientist \cite{lu_ai_2024} that employs an iterative process in which an LLM generates queries from a research idea to retrieve relevant papers via the Semantic Scholar API \cite{kinney_semantic_2023}. The LLM then compares the idea against these papers until a clear decision is reached or a preset iteration limit is met. However, this approach has several limitations. First, it depends on keyword-based retrieval methods to get the most relevant papers to an idea, which may fail if the relevant papers do not contain the exact keywords. Second, comparing an idea against a large number of retrieved papers (sometimes over 100) can introduce known issues that LLMs often overlook instructions within a prompt \cite{loya_exploring_2023, sclar_quantifying_2023, joshi_coprompter_2025}. Finally, the decision of novelty evaluation relies on string matching for phrases like "decision made: novel" or "decision made: not novel." If such a decision is not reached, the idea is automatically considered novel. Independent evaluations \cite{beel_evaluating_2025} have further highlighted challenges in AI Scientist’s novelty assessments, noting that the system can misclassify well-established concepts (micro-batching for stochastic gradient descent) as novel.

Our work builds on these insights by combining retrieval-then-rerank methods \cite{zhou_towards_2022, naik_literature-augmented_2022}  and uses expert-annotated examples to ensure that our novelty evaluations are grounded in the relevant literature.

%% file: paper/formative.tex
\definecolor{lilac}{RGB}{220, 200, 220} 

\section{Formative Study on Challenges in Evaluating Novelty}

Evaluating idea novelty in scientific literature is inherently challenging because the criteria for novelty are subjective and can be defined in multiple ways. We conducted a formative study, referred to as the expert-annotated study throughout the paper, where the first and second authors reviewed the novelty of ideas based on the most relevant papers.

To assess idea novelty relative to existing literature, our study engaged experts who evaluated 51 ideas, comprising of 46 generated by the Scideator system \cite{radensky_scideator_2024} and 5 adapted from accepted and rejected papers from OpenReview (ICLR 22, NeurIPS 23).\footnote{Fewer examples were taken from OpenReview since the primary focus was on evaluating ideas from Scideator.} Each idea was classified into one of three categories: novel, moderately novel, or not novel. For every idea, we identified the most relevant papers through a two-step process: candidate papers were initially gathered using keyword-based queries and subsequently re-ranked using an LLM-based reranker \cite{sun_is_2023} according to their overall relevance to the idea.

The experts achieved a moderate agreement (Cohen's Kappa = 0.64). A key challenge identified was that experts sometimes relied on their broader domain knowledge rather than restricting their judgments to the most relevant papers, as the top papers alone were often not sufficient. Additionally, using three categories led to disagreements, as the distinction between novel and moderate novelty is itself subjective.

Building on these observations, we conducted a second study to minimize the influence of external knowledge. In this study, experts were instructed to base their judgments solely on the provided papers, and the categories were simplified to just two: novel and not novel.

Inspired by prior work \cite{portenoy_bursting_2022,kang_augmenting_2022,chan_31_2018, suh_luminate_2024,srinivasan_improving_2024,choi_creativeconnect_2024,kang_biospark_2024, radensky_scideator_2024}  that categorizes research ideas into core facets such as purpose (the problem being addressed by the paper)  and mechanism (the proposed solution to the problem), we define novelty as follows: An idea is considered novel if it differs from all retrieved papers in at least one core facet for the topic at hand—namely, purpose (i.e., a distinct objective), mechanism (i.e., a distinct technical approach), or evaluation (i.e., a distinct validation method). An idea is also considered novel if it uniquely combines these facets or applies them to a new application domain.

Using this controlled framework, we reannotated a set of ideas and evaluated 51 ideas, comprising of 34 new ones generated by \citet{radensky_scideator_2024} and 17 from the previous study where external knowledge had influenced novelty judgments. By narrowing the focus to the relevant papers alone, we observed fewer disagreements and achieved a higher agreement rate (Cohen's Kappa = 0.68). Of the 8 instances of disagreement, in 4 cases one expert overlooked details from the paper, in 2 cases the experts differed in their perception of subtle contributions to novelty, and in the remaining 2 cases no specific comments were provided. 

This formative study highlights that a robust novelty checker depends critically on high-quality retrieval and a well-defined notion of novelty. These findings directly inform our methodology described in the following section.

%% file: paper/approach.tex
\section{Methodology: Idea Novelty Checker}

\begin{figure*}[t]
    \centering
    \resizebox{\textwidth}{!}{\includegraphics{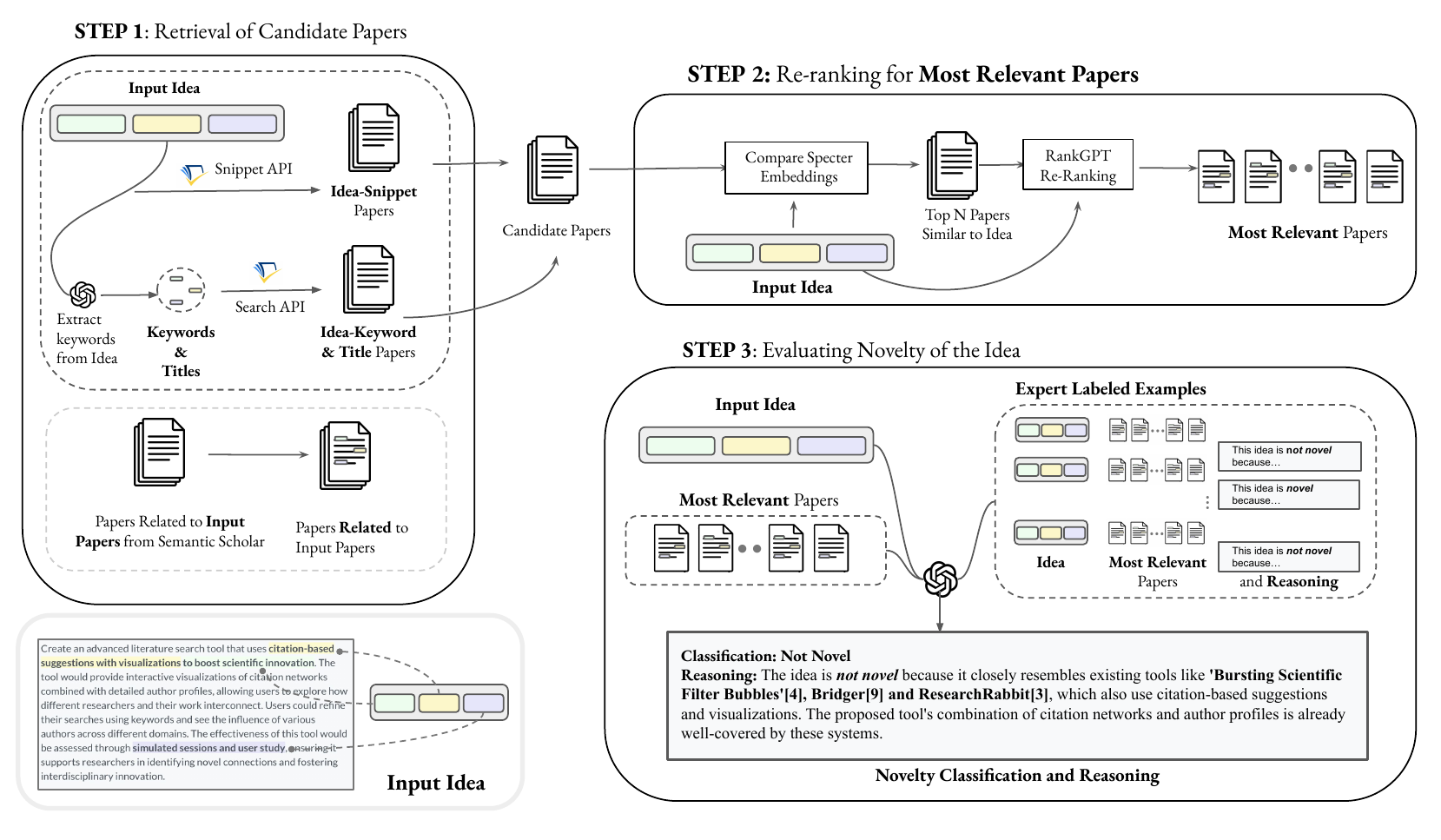}}
    \caption{Our \method\ follows a retrieve-then-rerank approach for novelty evaluation. First, it gathers a broad set of papers relevant to an idea using query expansion (extracting keywords and titles from the idea) and snippet search (using the entire idea as input). Optionally, if seed papers are provided, we retrieve papers similar to these seed papers. Next, a two-stage re-ranking process is applied, where an embedding-based ranking strategy filters the large collection to top-$N$ papers, followed by a facet-based LLM re-ranker to identify the top-$k$ most relevant papers. Finally, these top-$k$ papers are used to assess the idea's novelty, guided by in-context examples that evaluate novelty with grounded reasoning.}
    \label{fig:novelty_checker}
\end{figure*}

Based on our formative study findings, our novelty checker is designed with two key components that address the two critical challenges: \colorbox{lilac}{C1} ensuring high-quality retrieval of papers relevant to the idea for novelty assessment and \colorbox{lilac}{C2} establishing clear criteria for judging novelty. The challenge \colorbox{lilac}{C1} arises from the vast space of overlapping papers—there are hundreds of millions of potential matches. To address this, our system first filters the scientific literature to collect the most relevant papers for a given idea (see Section \ref{sec:approach_mrp} and Step 1 and 2 in Figure \ref{fig:novelty_checker}). The input idea is then compared to each paper in this collection by prompting an LLM (see Section \ref{sec:approach_noveltycheck}).

In tackling challenge \colorbox{lilac}{C2}, which arises from the inherent subjectivity and multiple definitions of novelty, the novelty checker leverages expert-labeled examples of \textit{novel} and \textit{not novel} ideas from the formative study. It generates reasoning grounded not only in comparisons against the most relevant papers but also in the standardized definition of novelty introduced earlier, which helps counteract subjectivity (see Step 3 in Figure \ref{fig:novelty_checker}). Below, we detail these two components.

\subsection{Most Relevant Papers to Idea}\label{sec:approach_mrp}

Following established information retrieval practices 
\cite{gao_llm4rerank_2025, abdallah_rankify_2025, nouriinanloo_re-ranking_2024, meng_ranked_2024, sun_is_2023, goharian_twolar_2024}, our pipeline uses a two-phase approach for identifying the most relevant papers to a given idea. First, we gather a broad set of candidate papers related to the idea. Then, we re-rank these candidates in two steps: first using embedding-based similarity, and then applying LLM-based re-ranking to facet-based similarity.

\subsubsection*{\textsc{STEP 1}: Retrieval of Candidate Papers} 
To accurately assess the novelty of an idea, it is crucial to compare it against a comprehensive collection of papers that cover the various facets of the idea. For a given idea and its corresponding papers (if any) used to generate the idea, we find more related papers to these input seed papers using the Semantic Scholar API \footnote{\href{https://api.semanticscholar.org/api-docs/recommendations\#tag/Paper-Recommendations/operation/get\_papers\_for\_paper}{api.semanticscholar.org/api-docs/recommendations}}. However, simple retrieval methods often overlook important aspects of an idea \cite{mysore_csfcube_2021, mysore_multi-vector_2022, wang_doris-mae_2023}.  To improve the paper collection's coverage we follow \cite{lu_ai_2024, si_can_2024} and employ a query-based retrieval method, where search queries are generated corresponding to different keywords related to the idea, and queried through the Semantic Scholar Search API \cite{kinney_semantic_2023}. Corresponding to each search query, papers are added to the collection of relevant papers. We prompt the LLM ($LLM_{query}$) to generate these search queries based on the keywords and potential titles related to the idea.  

Next, we also employ Semantic Scholar's snippet search\footnote{\href{https://api.semanticscholar.org/api-docs/\#tag/Snippet-Text}{api.semanticscholar.org/api-docs/\#tag/Snippet-Text}}, which is trained to identify similar snippets (approximately 500 words of text) in other papers. We leverage the context size of this retrieval mechanism by incorporating the entire idea into the snippet search. Finally, we combine the seed papers and their related works with the papers retrieved from the two Semantic Scholar based query-retrieval method. This combined set form the candidate papers for the ideation process.

\vspace{-2mm}
\subsubsection*{\textsc{STEP 2}: Re-ranking for Most Relevant Papers} 
\vspace{-1mm}
To identify the papers most likely to overlap with the candidate idea, we implement a two-stage re-ranking process that combines embedding-based filtering with an LLM-based re-ranking approach. To identify the papers most likely to overlap with the candidate idea, we implement a two-stage re-ranking process that combines embedding-based filtering with an LLM-based re-ranking approach. 

First, we employ \textbf{embedding-based filtering} to compute the semantic similarity between the idea and each paper in our collection of papers from \textsc{STEP 1}. We select the top $N$ papers with the highest cosine similarity between their embeddings and the idea embedding. While this embedding-based ranking efficiently narrows down the collection of papers, it is limited in its capacity to capture deeper and more contextual relationships between different facets of the idea and the papers, in comparison to powerful state-of-art LLMs~\cite{reimers_sentence-bert_2019}.

To address these limitations we employ a popularly used \textbf{LLM-based re-ranker}, RankGPT \cite{sun_is_2023}, which refines the initial ranking of candidate papers by examining how relevant each paper is to the idea. We change relevance criteria to match it with each key facet of the idea. RankGPT goes beyond simple surface similarities by comparing the papers against the idea’s application domain, purpose, mechanism, and evaluation. It follows a clear set of priorities: first, it favors papers that match all key facets of the idea; then, it prefers those that align with the application domain and purpose; next, it considers papers that share similarities in purpose, mechanism, or evaluation; and finally, it ranks lower those that only partially match or address related facets. This approach ensures that the final ranking accurately reflects the relevance and depth of each paper in connection with the idea. We refer to the LLM used for RankGPT as ($LLM_{rankgpt}$).

\noindent This collection of $k$-most relevant papers is used by the novelty checker in the next step to evaluate the idea's novelty.

\subsection{Idea Novelty Evaluation}\label{sec:approach_noveltycheck}

To assess an idea's novelty, we prompt an LLM ($LLM_{novelty}$) with both the idea and its top-$k$ relevant papers. The LLM outputs a binary classification (novel or not novel) accompanied by reasoning based on the top-$k$ retrieved literature. To guide the LLM's judgment, we include $n_{examples}$ in-context examples drawn from our formative study, where $n_{examples}$ is treated as a hyperparameter. These examples reflect the experts' criteria for novelty: an idea is considered novel if it differs from all retrieved papers in at least one core facet—namely, purpose (i.e., objective of idea), mechanism (i.e., technical approach), evaluation (i.e., validation method), a unique combination of these facets, or if it applies the same facets to a new application domain.

%% file: paper/implementation.tex
\section{Implementation \& Baselines}

\paragraph{\textbf{Dataset:}} From our formative study, we collected 67 consensus-labeled examples (39 labeled as novel and 28 as non-novel). We split into training and test sets (35 for training and 32 for testing) with a balanced distribution of novel and non-novel ideas.  Please refer to Table \ref{table:expert-labeled-examples} in the Appendix for sample examples.

\paragraph{\textbf{Baselines:}} We evaluated multiple baselines to benchmark our novelty assessment approach. First, we employed a zero-shot prompt as a straightforward baseline, and further refined this manually written prompt using Anthropic's prompt generator\footnote{\url{https://docs.anthropic.com/en/docs/build-with-claude/prompt-engineering/prompt-generator}}. We also applied popular prompt optimization techniques such as DSPy \cite{khattab_dspy_2023} and TextGRAD \cite{yuksekgonul_textgrad_2024}, which optimize the prompt instructions using a train/validation split created from formative study examples. 

As an alternative to using in-context examples from the formative study, we extracted reviews from ICLR and NeurIPS submissions via the \href{https://openreview.net}{OpenReview} API. These reviews comprise aspects such as \textit{strengths, presentation, limitations, soundness, weaknesses, questions, confidence, contribution, summary,} and \textit{rating}. The input title and abstract were adapted to match the ideas in the training data using a style-change prompt\footnote{All prompts are provided in the anonymised codebase.}. After rigorous filtering, we identified approximately 8,156 submissions discussing idea novelty and manually selected reviews that specifically evaluated the core idea rather than the entire paper. From these, we randomly sampled 20 idea-review pairs to serve as an additional baseline with different in-context examples.

In addition to these baselines, we also compare our novelty checker `\textit{prompt}' with that of AI Scientist \cite{lu_ai_2024} (different from its paper reviewer) and AI Researcher \cite{si_can_2024} on the same test set of ideas and fixed top 10 papers. We compare \textbf{only the prompts to assess novelty} of these two approaches with ours, rather than the entire system, because the test set containing the novelty judgments by experts were based on a fixed set of the 10 most relevant papers for each idea. Since different retrieval methods could introduce new papers and potentially change novelty classification, we standardize the most relevant papers to ensure a fair comparison of the prompts alone. Additionally, since both setups require a different style of input idea, we adapted the ideas to match the requirements of each system.
 

\begin{table*}[tp!]
\centering
\small

\centering
\begin{tabular}{lccccc}
\toprule
Models & Accuracy & Precision & Recall & F1 & Cohen Kappa \\ 
\midrule
\rowcolor[HTML]{EFEFEF} 
\textbf{Zero Shot Setting} & & & && \\
Zero Shot  & 0.68 & 0.76 & 0.64 & 0.65 & - \\
+ improved prompt using  Anthropic prompt generator & 0.68 & 0.70 & 0.64 & 0.64 &- \\
\midrule
\rowcolor[HTML]{EFEFEF} 
\textbf{Prompt Optimizers}    &    &     &    &  & \\
\textbf{DSPy}   &    &     &    &  & \\

- with idea, most relevant papers, class &  0.68 & 0.83 & 0.62 & 0.58& - \\

- with idea, most relevant papers, class, reasoning &  0.66 & 0.82 & 0.58 & 0.52&  \\

\textbf{TextGRAD}    &    &     &    &   \\ 

- with idea, most relevant papers, class &  0.78  & 0.76 & 0.76 & 0.76&- \\
\midrule
\rowcolor[HTML]{EFEFEF} 
\textbf{In-context Setting}   &    &     &    & &  \\
\textbf{Open-Review Examples} &    &     &    & &  \\
- with idea \& review (i.e.,~reasoning) & 0.59 & 0.55 & 0.51 & 0.43 &- \\
\textbf{Expert Labeled Examples} & & & && \\

- with idea, reasoning & 0.75 & 0.76 & 0.77 & 0.75&- \\

- with idea, most relevant papers, class &  0.78  &  0.77   & 0.76   & 0.77 &- \\
- with idea, most relevant papers, class, reasoning & \underline{0.81} & \underline{0.84} & \underline{0.78} & \underline{0.79}& \underline{0.59}\\
\midrule
\rowcolor[HTML]{EFEFEF} 
\textbf{Other Novelty Checkers}   &    &     &    & &  \\
\textbf{AI Scientist} \cite{lu_ai_2024} & 0.47 & 0.55 & 0.53 & 0.44 & 0.05  \\
\textbf{AI Researcher} \cite{si_can_2024} & & & & & \\
- \textsc{GPT-4o} & 0.78 & 0.81  &  0.74   & 0.75 & 0.52 \\
- \textsc{Claude-3-5-sonnet} & 0.56 & 0.63  &  0.61  & 0.56 & 0.19 \\
\bottomrule
\end{tabular}
\caption{Experimental Results using \textsf{gpt-4o} on expert-annotated dataset.}
\label{table:novelty-experiments}
\vspace{-3mm}
\end{table*}

\paragraph{\textbf{Implementation Settings:}} 
For our novelty evaluation system, we use SPECTER-2 \cite{cohan_specter_2020} as the default embedding model. Initially, we retrieve the top $N=$100 papers using these embeddings, from which the top $k=$10 most relevant papers are selected for comparison with the input idea. The default language model for the idea keyword extraction ($LLM_{query}$), re-ranking process ($LLM_{rankgpt}$), and novelty evaluation ($LLM_{novelty}$) is \textsf{\small{gpt-4o}}\footnote{We used the model "gpt-4o" during August and September 2024.} . Expert-labeled data from the formative study is incorporated as in-context examples in the novelty checker. We experimented with various numbers of in-context examples (comprising idea-paper pairs along with their novelty class and reviews) and found that the best performance was achieved using 15 idea examples (random seed 100). For the OpenReview examples, the best setup involved 5 idea-review pairs. For DSPy we used 2 bootstrapped examples, and trained both DSPy and TextGRAD for 12 prompt iterations. 

%% file: paper/experiments.tex
\section{Experiments}

In this section, we first compare different baselines on the dataset for novelty evaluations (Section \ref{sec:experiment_baseline}). Next we present our findings from ablation studies that shows the imporance of each component in our approach (Section \ref{sec:experiment_ablation}). ablations studies  We supplement these findings with qualitative examples of expert-labeled ideas and compare our setup with recent novelty checkers (Section \ref{sec:qualitative_analysis}). We conclude with insights from prompt optimization experiments that highlights the sensitivity of LLMs to prompt variations for novelty evaluation tasks (Section \ref{sec:nc_prompt_sensitivity}). 

\subsection{Comparing Novelty Checker Prompts}
\label{sec:experiment_baseline}
Our experiments show that incorporating expert-annotated data as in-context examples significantly enhances novelty classification accuracy compared to zero-shot prompts, DSPY, TextGRAD, and setups using OpenReview examples (Table \ref{table:novelty-experiments}).
 Since OpenReview reviews do not reference the associated papers, we evaluated our expert-labeled examples both with and without including relevant papers to ensure a fair comparison. Notably, even when we excluded the relevant papers from the expert-labeled examples, our approach still outperformed the OpenReview baseline.

Additionally, we compared two configurations for DSPY, one with reasoning and one without. Our expert-labeled prompt consistently achieved higher performance than the prompt optimizations produced by these methods, and we posit that the number of examples for train/validation were not sufficient for prompt optimisers with gpt-4o. The TextGRAD prompt optimiser did not improve upon its initial system prompt.  It provided valuable insights into the LLM's prompt sensitivity, which we further discuss in Section \ref{sec:nc_prompt_sensitivity}.

Our approach achieved over \textbf{10 times more agreement with expert-labeled examples} compared to AI Scientist, and \textbf{approximately 13\% higher agreement} than AI Researcher, further validating the effectiveness of our novelty checker. It is important to note that AI Scientist defaults to "not novel" when it fails to reach a conclusion in novelty evaluation (18 out of 32 times), which may have impacted its agreement rates. We also present some qualitative examples in Figures \ref{fig:comparison_reviews_ideas}, \ref{fig:comparison_reviews_ideas_2} and \ref{fig:comparison_reviews_ideas_3} of the Appendix, showcasing how these approaches evaluate the novelty of an idea.

%% file: paper/ablation.tex
\subsection{Ablation Studies}
\label{sec:experiment_ablation}
\textbf{Setup:} To assess the contribution of each component in our novelty checker, we conducted ablation studies using 58 ideas (comprising 13 `not novel' instances from our test set and 45 NLP papers from the literature). For this experiment, we focus on the `not novel' cases, since the ideas labeled novel in expert-labeled test data can vary with different retrieved paper sets.  In our ablations, we considered the following variations: (i) \textbf{Complete System:}  Uses both keyword and snippet retrieval (each returning the top-\(k\) documents based on Semantic Scholar's ranking), embedding filtering, and facet-based RankGPT re-ranking; (ii) \textbf{RankGPT Relevance:} Used the same retrieval methods (keyword and snippet) plus embedding filtering, but replaced the facet-based RankGPT re-ranker with one based on general relevance \cite{sun_is_2023}. This variation differs from the complete system only in the LLM re-ranking component, allowing us to assess the importance of facet-based re-ranking; (iii) \textbf{Embedding Filtering:} Omits the LLM re-ranker entirely, relying only on the embedding-based filtering. This setup allows us to assess the importance of the LLM re-ranking step; and (iv) \textbf{Snippet Retrieval} and \textbf{Keyword Retrieval}: Each of these setups returned the top-\(k\) documents from their respective retrieval method (without embedding filtering or any LLM re-ranking), leveraging the inherent ranking/scoring provided by Semantic Scholar. This setup allows to assess the importance of both re-ranking steps. This structured setup enabled us to isolate the contribution of each component (retrieval method vs. re-ranking strategy) and evaluate whether they collectively brought key papers for novelty assessment into the top 10. We use \textsf{o3-mini} for evaluating novelty (Step 3) and \textsf{gpt-4o} for re-ranking (Step 2).

\noindent\textbf{Classification Analysis:}  Table~\ref{tab:class_ablation} shows that the complete system, which employs facet-based re-ranking in RankGPT, significantly outperforms its ablated variants in accuracy.
The results demonstrate that methods relying only on keyword or snippet-based retrieval have much lower accuracy, and even alternate re-ranking strategies with a single embedding-based reranker or both embedding and general relevance RankGPT are insufficient to consistently bring key papers into the most relevant paper set. These findings show that combining facet-based reranking with embedding is critical  for identifying the most relevant papers.

\begin{table}[ht]
\centering
\vspace{-2mm}
\caption{Accuracy of predicting ``not novel''.}
\label{tab:class_ablation}
\small
\begin{tabular}{lc}
\toprule
\textbf{Method} & \textbf{Accuracy} \\
\midrule
Complete System           & 89.66\% \\
- Relevance RankGPT     & 13.79\% \\
- Embedding Filtering   & 10.34\% \\
- Snippet Retrieval     & 8.62\%  \\
- Keyword Retrieval     & 5.17\%  \\
\bottomrule
\end{tabular}
\vspace{-3mm}
\end{table}

\noindent\textbf{Analysis of the Most Relevant Papers:} Table~\ref{tab:doc_ablation} compares the top-10 most relevant papers retrieved under each ablation setting with those from the complete system. Approximately 30\% of the papers differ when using either embedding-based or general relevance RankGPT. Additionally, notable rank shifts are observed between the facet-based and relevance-based LLM rerankers. In contrast, without the reranking steps, both snippet and keyword retrieval exhibit minimal overlap with the final system’s top results, highlighting the importance  of the reranker stage. 

\begin{table}[ht]
\vspace{-3mm}
\caption{Comparing rank and overlap in retrieved papers with each variant to the complete system.
\emph{Overlap} indicates how many papers overlap on average with the complete system top-10 papers. 
\emph{Rank Shift} measures the average absolute difference in rank positions (only among overlapping papers).}
\label{tab:doc_ablation}
\resizebox{\columnwidth}{!}{%
\begin{tabular}{lcc}
\toprule
\textbf{Method} & \textbf{Overlap} ($\uparrow$) & \textbf{Rank Shift} ($\downarrow$) \\
\midrule
Relevance RankGPT     & 7.97  & 0.67  \\
Embedding Filtering & 7.93  & 0.84 \\
Snippet Retrieval     & 2.88  & 1.85 \\
Keyword Retrieval     & 1.17  & 1.39  \\
\bottomrule
\end{tabular}
}
\vspace{-5mm}
\end{table}

%% file: paper/qualitative.tex
\subsection{Qualitative Analysis} \label{sec:qualitative_analysis}



Table \ref{table:expert-labeled-examples} in the Appendix shows examples from our training set, including an idea, its most relevant papers, and the corresponding expert reasoning. While assessing novelty, we add both the titles and abstracts of the most relevant papers for each idea.

Figures \ref{fig:comparison_reviews_ideas}, \ref{fig:comparison_reviews_ideas_2}, and \ref{fig:comparison_reviews_ideas_3} in the Appendix qualitatively compare novelty evaluations by AI Scientist, AI Researcher, and \method~(ours) on two research ideas. \method\ provides concise justifications for its novelty decisions by referencing key similarities and differences with existing works. For example, in Example 1, it correctly identifies the idea as `novel' by highlighting these aspects. In contrast, AI Researcher evaluates each paper individually, classfying an idea as `not novel' if any paper is considered citable; but in our examples, none of the papers were flagged as citable despite sharing similar purposes, leading to a `novel' classification.  Due to space constraints, we show insights only from the first paper for each example. Figure \ref{fig:comparison_reviews_ideas_3} indicates that while AI Scientist’s judgments generally align with the ground truth and offer actionable suggestions, it sometimes misinterprets the idea—as in Figure \ref{fig:comparison_reviews_ideas_2}, where its focus shifts from the idea to the accompanying code.

%% file: paper/prompt_sensitivity.tex
\subsection{Prompt Sensitivity} \label{sec:nc_prompt_sensitivity}

In our experiments with TextGrad, we investigated how specific prompt instructions influence an LLM’s ability to classify the novelty of an idea. Figures in Appendices \ref{fig:prompt_textgrad1}, \ref{fig:prompt_textgrad2}, and \ref{fig:prompt_textgrad3} present the accuracy of various prompts optimized with TextGrad on our dataset  (train=25, validation = 10, test = 32).

Prompts with both non-zero and zero validation accuracy included various instructions for evaluating the novelty of ideas, such as assessing the uniqueness of methods and their comparison to existing research.  Through this prompt optimization process, we observed interesting ways in which LLMs may evaluate novelty, like considering historical context, frequency of similar studies, comparative analysis with existing works, examining arguments for both novel and non-novel perspectives. However, prompts without these specific instructions also influenced accuracy, suggesting the complexity of novelty evaluation with LLMs.

Notably, some prompts with similar instructions showed different performance on validation data. For example, both prompt 3 (accuracy = 0) and prompt 9 (accuracy = 0.6) include instructions to evaluate if the idea introduces unique methodologies, and how it compares to existing work. However, the difference in their performance suggests that subtle variations in wording and instruction framing can significantly impact the classification performance. It remains unclear why certain prompts perform better despite having similar instructions.

Our analysis highlights the LLM's sensitivity to prompt design when assessing novelty of an idea. Even minor variations in wording and structure can lead to substantial performance changes, emphasizing the need for careful prompt engineering and well-chosen in-context examples to guide the LLM for idea novelty evaluation.

\vspace{-6mm}

\vspace{0.5cm}

%% file: paper/conclusion.tex
\section{Conclusion}

In this work, we propose \method\, a retrieval-augmented pipeline for evaluating the novelty of scientific ideas and generating literature-grounded rationales. Our formative study highlighted two main challenges in evaluating novelty: (1) retrieving the most relevant papers from a vast corpus, and (2) establishing a fixed notion of novelty due to its inherent subjectivity. To address the latter, we incorporate expert-annotated examples in our novelty checker where we consider an idea to be novel within a given topic domain if it (1) differs from all retrieved papers in at least one core facet—namely, purpose (a new objective), mechanism (a distinct technical approach), or evaluation (a distinct validation method); (2) uniquely combines these facets; or (3) applies them to a new application domain. 

Our experiments on an expert-annotated dataset demonstrate that \method\ outperforms two well-known recent baselines, and our ablation studies confirm the importance of each component in our system. Furthermore, qualitative comparisons and analyses of prompt sensitivity provide additional insights into novelty evaluation.

\section{Limitations \& Future Work}

While \method\ is superior in many aspects, it also has some limitations. For instance, due to context size constraints (with fifteen in-context examples for both \textit{novel} and \textit{not novel} categories), our analysis is restricted to the top 10 retrieved papers, which may disproportionately influence the overall novelty assessment. Additionally, our definition of novelty relies on expert annotations, and the same annotators who provided the in-context examples also classified the test ideas. This could potentially give our approach an advantage in understanding our view of novelty. Moreover, many of the ideas used for testing were generated by the same system \cite{radensky_scideator_2024} that produced the in-context examples, although some ideas were sourced from OpenReview. 

In future work, we aim to address these limitations by expanding the literature scope using tools such as \href{https://openai.com/index/introducing-deep-research/}{OpenAI Deep Research} and ScholarQA \cite{singh_ai2_2025} and further refining our novelty evaluation to view novelty as a continuum rather than binary classification.

%% file: paper/appendix.tex
\begin{table*}[htp!]
\vspace{-2mm}
\caption{\centering\textbf{Expert-labeled examples from annotation study}}\label{table:expert-labeled-examples}
\vspace{-6mm}
\resizebox{\textwidth}{!}{ 
\begin{tabular}{p{15cm}}
 \\
\hline
\rowcolor[HTML]{EFEFEF} 
\textbf{Example 1} \\
\hline

\textbf{Idea:} Develop a \textbf{natural language processing classifier designed to improve scientific paper revisions} by automatically identifying and categorizing reviewer comments that are most likely to lead to substantial and actionable revisions. The system would be trained on a \textbf{manually-labeled dataset analysis} of scientific review comments and the corresponding paper edits, leveraging features such as linguistic cues, sentiment, and comment specificity to predict the likelihood of a comment being acted upon. This classifier could then be used to prioritize reviewer feedback, helping authors focus on the most impactful suggestions first. \\
\hline
\textbf{Most Relevant Papers:}
\vspace{-2mm}
\begin{enumerate}
    \setlength{\itemsep}{0pt} 
    \setlength{\parskip}{0pt} 
    \item \href{https://www.semanticscholar.org/paper/fd8c64d0b912795e1cefc0aba4c6d90499132755}{ARIES: A Corpus of Scientific Paper Edits Made in Response to Peer Reviews}
    \item \href{https://www.semanticscholar.org/paper/f2209eb5ac6747319a29b87dedabb97770be3243}{Can large language models provide useful feedback on research papers?}
    \item \href{https://www.semanticscholar.org/paper/53489041a08ef1a6cd19b4c95ce092a148283b6b}{A Dataset of Peer Reviews (PeerRead): Collection, Insights and NLP Applications}
    \item \href{https://www.semanticscholar.org/paper/e53b0d9c061a1a108d87d79826c47cf77bac85d6}{arXivEdits: Understanding the Human Revision Process in Scientific Writing}
    \item \href{https://www.semanticscholar.org/paper/7c311d7918800fc842dd50a46e6e9a8df86d6424}{Characterizing Text Revisions to Better Support Collaborative}
    \item \href{https://www.semanticscholar.org/paper/cefd3993db4d065b95ab8f105452fb728c02b60e}{Can We Automate Scientific Reviewing?}
    \item \href{https://www.semanticscholar.org/paper/28c26e7b3a30fa2808ed103aade1fcb4d752906e}{DeepReviewer: Collaborative Grammar \& Innovation Neural Network for Paper Review}
    \item \href{https://www.semanticscholar.org/paper/2b4edb9515a26561ea3f9ee2a63a506721c8369e}{Aspect-based Sentiment Analysis of Scientific Reviews}
    \item \href{https://www.semanticscholar.org/paper/a091744cbcf4e209f1accf098671d0810282e7bf}{Aspect-based sentiment analysis of online peer reviews and prediction of paper acceptance}
    \item \href{https://www.semanticscholar.org/paper/f14b82fd35335667083d1141f6d1fc65e614644e}{ReviVal: Towards Automatically Evaluating the Informativeness of Peer Reviews}
    \vspace{-2mm}
\end{enumerate}

\\
\hline
\textbf{Reasoning:} The idea is \textbf{novel} because it uniquely focuses on prioritizing reviewer comments for actionable revisions, which is not explicitly addressed in ARIES[1] or other works like ReviVal[10]. \\
\hline
\rowcolor[HTML]{EFEFEF} 
\textbf{Example 2} \\
\hline

\textbf{Idea:} Develop a \textbf{systematic review-based framework} designed \textbf{to align LLM evaluation with human preferences}, ensuring that evaluation criteria are continuously refined based on comprehensive reviews of user feedback and emerging model behaviors. This framework will utilize \textbf{content analysis of user interactions and feedback} to identify patterns and areas of improvement. The effectiveness of this framework will be assessed through a \textbf{qualitative study} involving iterative cycles of user feedback and criteria refinement. \\
\hline
\textbf{Most Relevant Papers:}
\vspace{-2mm}
\begin{enumerate}
\setlength{\itemsep}{0pt} 
    \setlength{\parskip}{0pt} 
    \item \href{https://www.semanticscholar.org/paper/a0d83f9e15e722f23c14eb83cb2f87c1d1ea6400}{EvalLM: Interactive Evaluation of Large Language Model Prompts on User-Defined Criteria}
    \item \href{https://www.semanticscholar.org/paper/78105fd895d648f80374e75c47857cd36edce3f2}{Humanely: Human evaluation of LLM yield, using a novel web-based evaluation tool}
    \item \href{https://www.semanticscholar.org/paper/7a4bbfb0fdddbaa20214b90bffe0a0fc1d1aedf8}{Evaluation of Code Generation for Simulating Participant Behavior in Experience Sampling Method by Iterative In-Context Learning of a Large Language Model}
    \item \href{https://www.semanticscholar.org/paper/2ab90f60ea2e0d345f8f6f66dd4efa5d347eac25}{Human-Centered Evaluation and Auditing of Language Models}
    \item \href{https://www.semanticscholar.org/paper/b288562f97295479d3f205fb118d276bb12966ca}{Aligning Model Evaluations with Human Preferences: Mitigating Token Count Bias in Language Model Assessments}
    \item \href{https://www.semanticscholar.org/paper/6098ac103ba222f9c3b089714ef3100357993255}{Who Validates the Validators? Aligning LLM-Assisted Evaluation of LLM Outputs with Human Preferences}
    \item \href{https://www.semanticscholar.org/paper/52570796ad1fdc20a367caf1d099e729d6456241}{Human-Centered Design Recommendations for LLM-as-a-judge}
    \item \href{https://www.semanticscholar.org/paper/afd83013ba5e6cefb0b1c09084e8c6a15a47e0c3}{CheckEval: Robust Evaluation Framework using Large Language Model via Checklist}
    \item \href{https://www.semanticscholar.org/paper/cef330bacf014d60daabbd489647b2006af130ca}{Discovering Language Model Behaviors with Model-Written Evaluations}
    \item \href{https://www.semanticscholar.org/paper/ecdd53eaab7455daea27609b07a418a21aa7ad35}{Prometheus 2: An Open Source Language Model Specialized in Evaluating Other Language Models}
    \vspace{-2mm}
\end{enumerate}

\\

\hline
\textbf{Reasoning:} The idea is \textbf{not novel} because it closely resembles existing frameworks like EvalLM[1] and HumanELY[2], which already align LLM evaluations with human preferences using user-defined criteria and human feedback. 
\\
\bottomrule
\end{tabular}
}
\end{table*}

\begin{figure*}[htp!]
    \centering
     \caption{Two example ideas used as the basis for comparison in subsequent figures, evaluated by \method\ (Ours), AI Scientist, and AI Researcher.}\includegraphics[width=\textwidth]{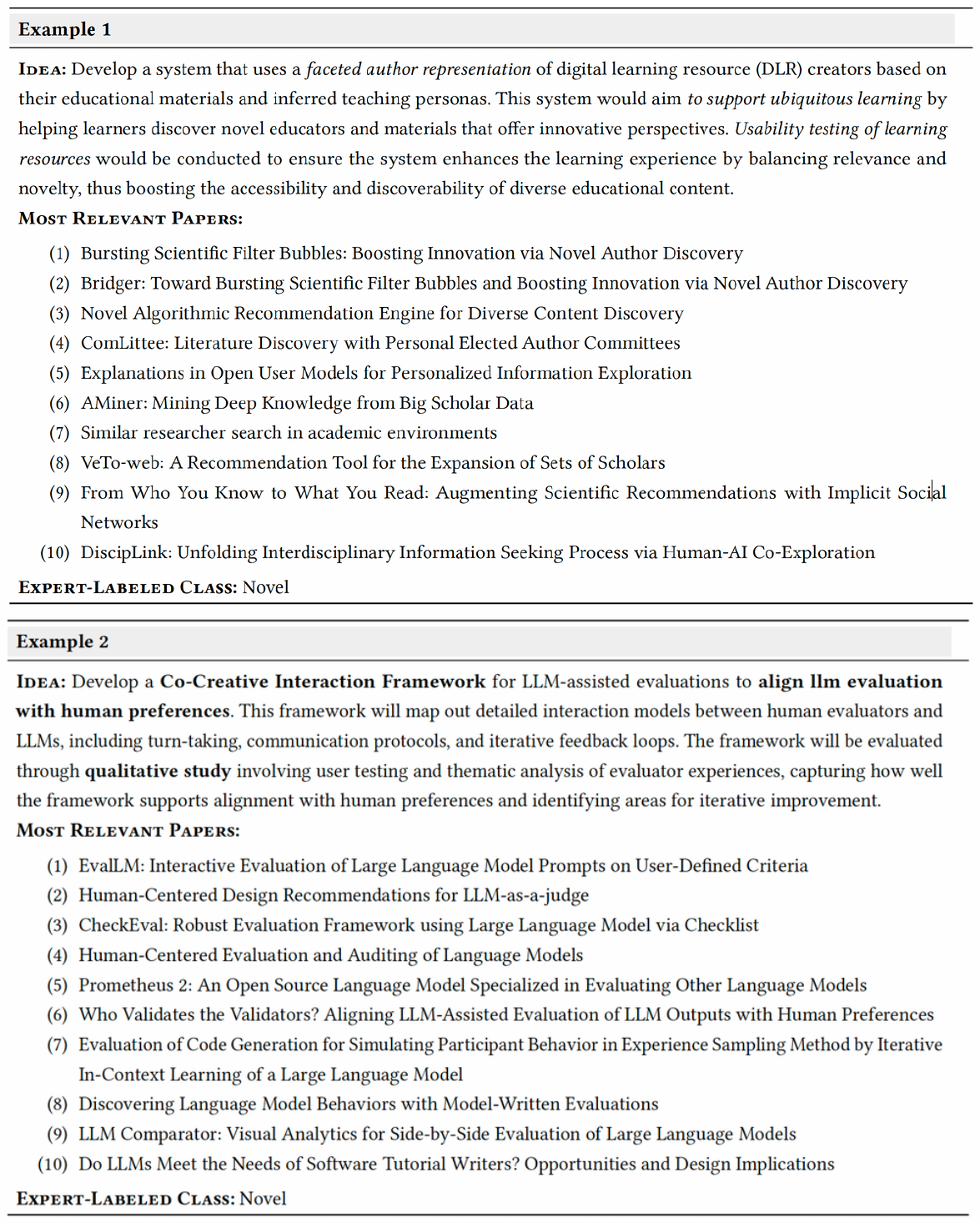} 
    \label{fig:comparison_reviews_ideas}
\end{figure*}

\begin{figure*}[htp!]
    \centering
    \caption{Reviews corresponding to idea 1 in Figure \ref{fig:comparison_reviews_ideas}.}\includegraphics[width=\textwidth]{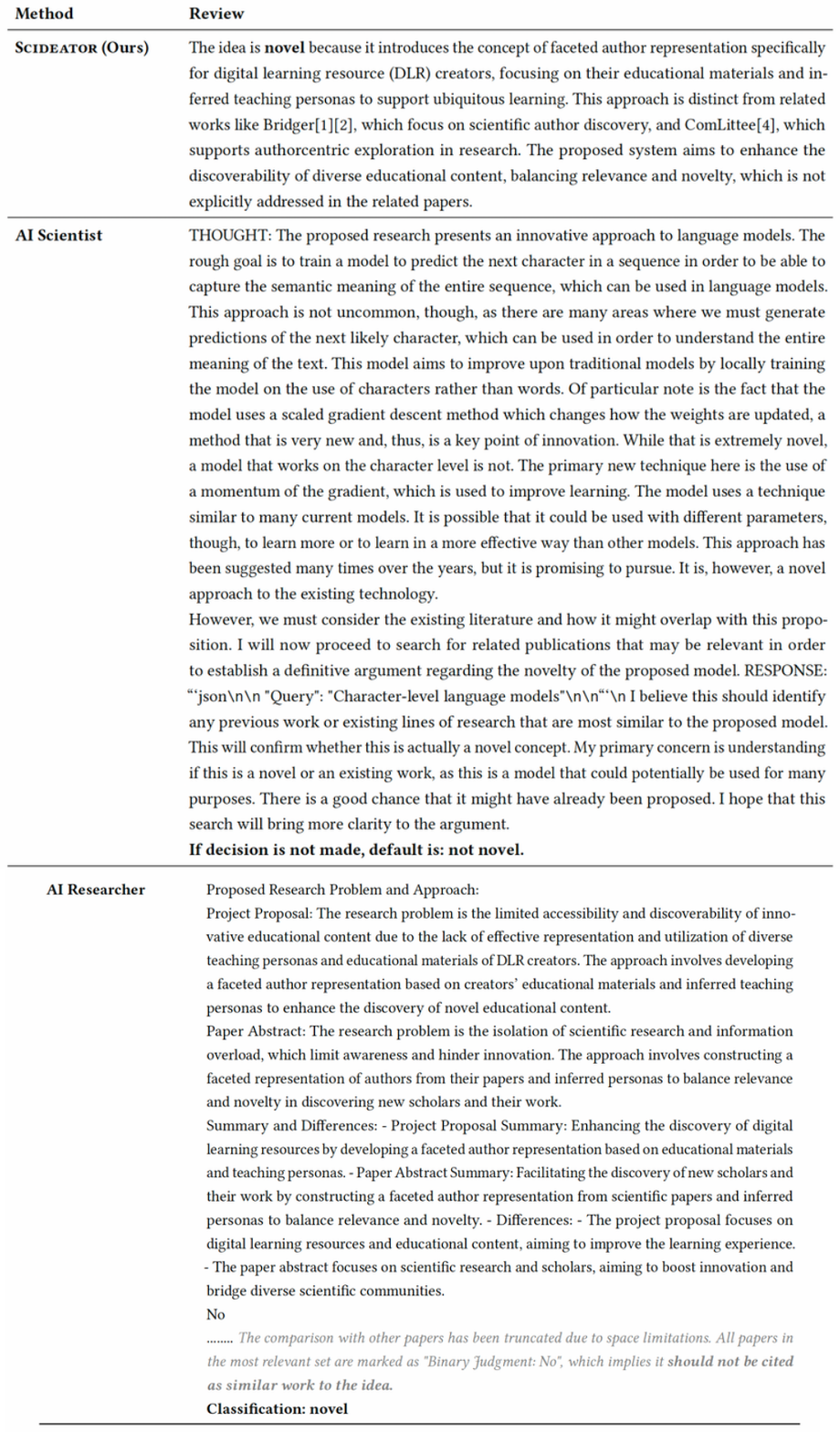} 
    \label{fig:comparison_reviews_ideas_2}
\end{figure*}

\begin{figure*}[htp!]
    \centering
    \caption{Reviews corresponding to idea 2 in Figure \ref{fig:comparison_reviews_ideas}.}\includegraphics[width=\textwidth]{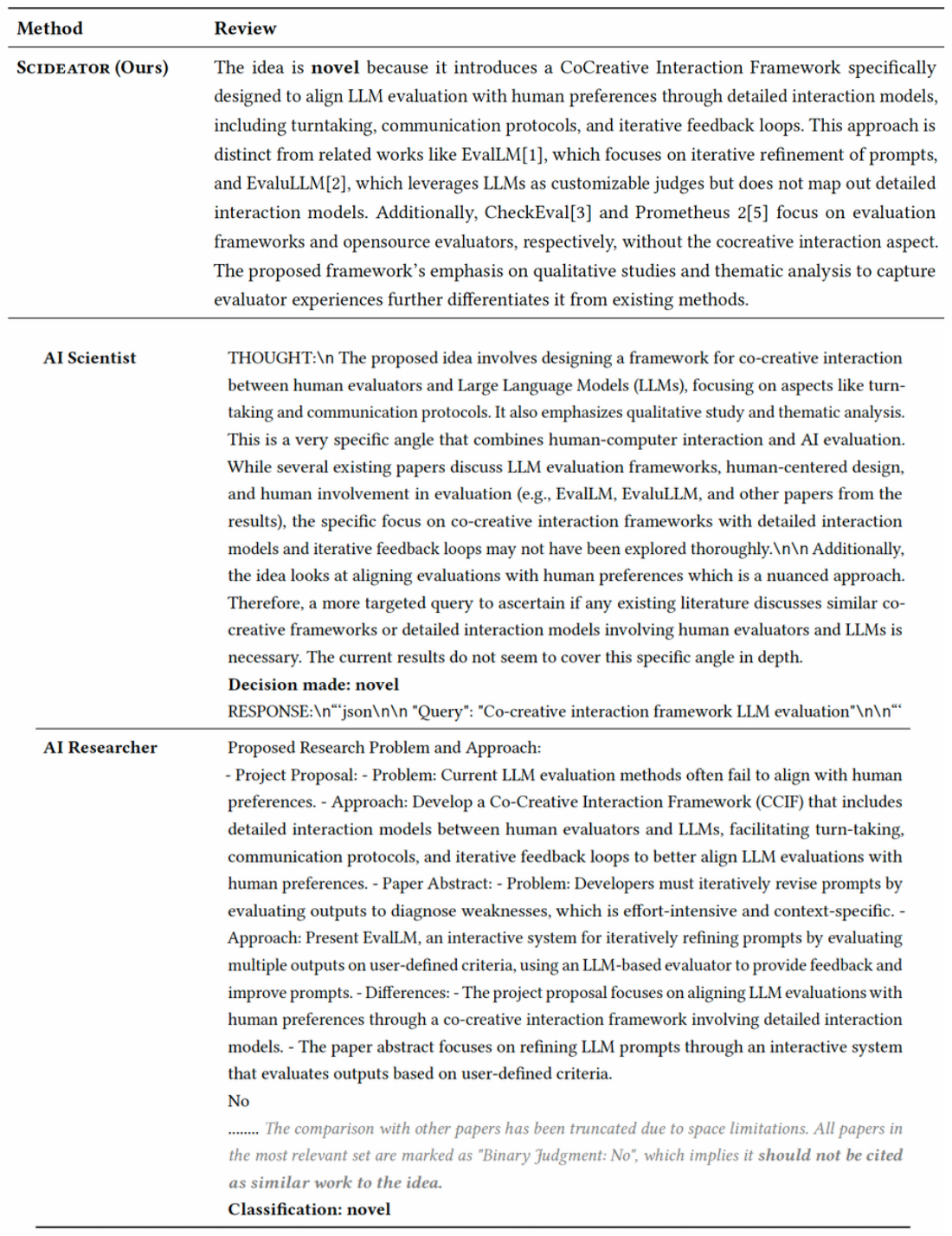} 
    
    \label{fig:comparison_reviews_ideas_3}
\end{figure*}

\begin{figure*}[htbp]
  \captionsetup{justification=centering}   
  \caption{\textbf{Performance trends of test accuracy across prompts during
    prompt optimization with TextGRAD.}\newline
    Highlighted text shows unique instructions used to evaluate the novelty of
    ideas. The final test accuracy was 0.78125, so none of the optimized prompts
    (1 – 12) out-performed the original.\vspace{-4mm}}
  \label{fig:prompt_textgrad1}             
  \centering
  \includegraphics[width=\textwidth]{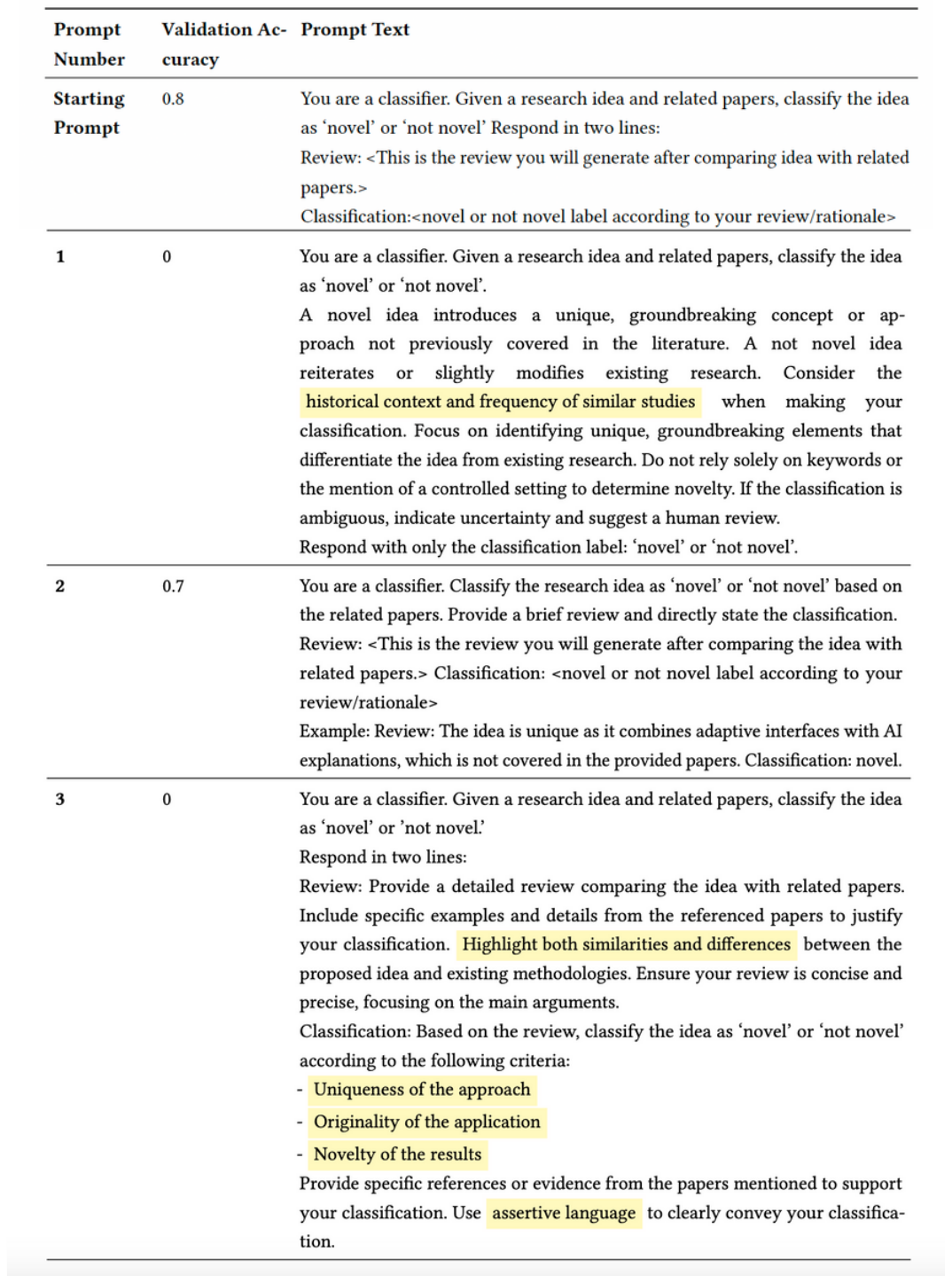}
  \vspace{-4mm}
\end{figure*}

\begin{figure*}[htp!]
    \centering
     \caption{contd. TextGrad Prompt Optimisation.}\includegraphics[width=\textwidth]{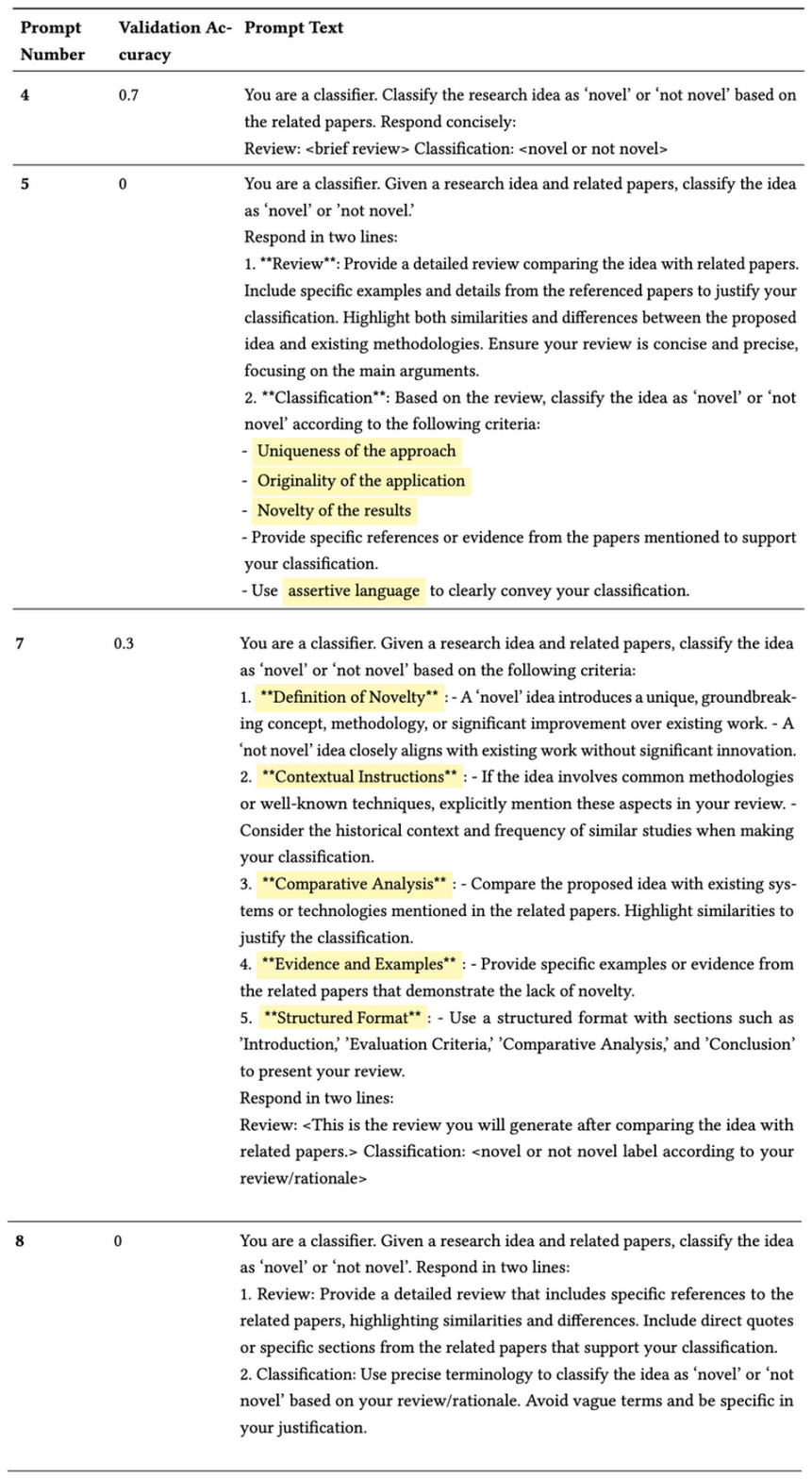}
   
    \label{fig:prompt_textgrad2}
\end{figure*}

\begin{figure*}[htp!]
    \centering
     \caption{contd. TextGrad Prompt Optimisation.}\includegraphics[width=\textwidth]{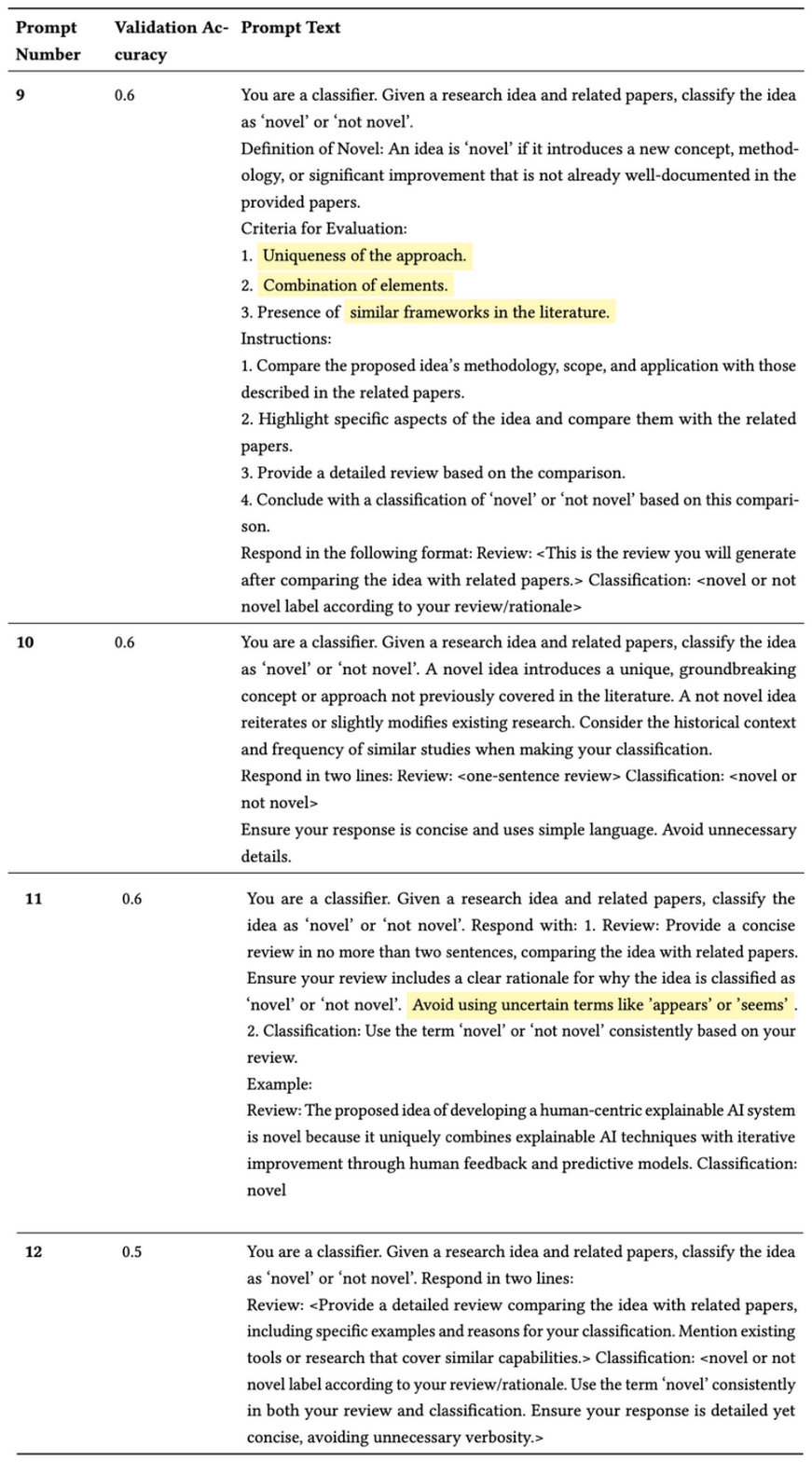}
    
    \label{fig:prompt_textgrad3}
\end{figure*}

%% file: acl_latex.bbl
\begin{thebibliography}{45}
\providecommand{\natexlab}[1]{#1}

\bibitem[{Abdallah et~al.()Abdallah, Piryani, Mozafari, Ali, and Jatowt}]{abdallah_rankify_2025}
Abdelrahman Abdallah, Bhawna Piryani, Jamshid Mozafari, Mohammed Ali, and Adam Jatowt.
\newblock \href {https://doi.org/10.48550/ARXIV.2502.02464} {Rankify: A comprehensive python toolkit for retrieval, re-ranking, and retrieval-augmented generation}.
\newblock Publisher: {arXiv} Version Number: 3.

\bibitem[{Ahmed et~al.()Ahmed, Ramachandran, Fuge, Hunter, and Miller}]{ahmed_interpreting_2019}
Faez Ahmed, Sharath~Kumar Ramachandran, Mark Fuge, Samuel Hunter, and Scarlett Miller.
\newblock \href {https://doi.org/10.1115/1.4041856} {Interpreting idea maps: Pairwise comparisons reveal what makes ideas novel}.
\newblock 141(2):021102.

\bibitem[{Baldelli et~al.()Baldelli, Jiang, Aizawa, and Torroni}]{goharian_twolar_2024}
Davide Baldelli, Junfeng Jiang, Akiko Aizawa, and Paolo Torroni.
\newblock \href {https://doi.org/10.1007/978-3-031-56027-9_29} {{TWOLAR}: A {TWO}-step {LLM}-augmented distillation method for passage reranking}.
\newblock 14608:470--485.
\newblock Book Title: Advances in Information Retrieval {ISBN}: 9783031560262 9783031560279 Place: Cham Publisher: Springer Nature Switzerland.

\bibitem[{Beel et~al.()Beel, Kan, and Baumgart}]{beel_evaluating_2025}
Joeran Beel, Min-Yen Kan, and Moritz Baumgart.
\newblock \href {https://doi.org/10.48550/ARXIV.2502.14297} {Evaluating sakana's {AI} scientist for autonomous research: Wishful thinking or an emerging reality towards 'artificial research intelligence' ({ARI})?}
\newblock Publisher: {arXiv} Version Number: 2.

\bibitem[{Bougie and Watanabe()}]{bougie_generative_2024}
Nicolas Bougie and Narimasa Watanabe.
\newblock \href {https://doi.org/10.48550/ARXIV.2412.10415} {Generative adversarial reviews: When {LLMs} become the critic}.
\newblock Publisher: {arXiv} Version Number: 1.

\bibitem[{Chan and Chang()}]{chan_31_2018}
Joel Chan and Joseph~Chee Chang.
\newblock \href {https://www.semanticscholar.org/paper/31-SOLVENT-%3A-A-Mixed-Initiative-System-for-Finding-Chan-Chang/de49225bd0d1beb7dc0bf91acd0a9a263c70176e} {31 {SOLVENT} : A mixed initiative system for finding analogies between research papers}.

\bibitem[{Choi et~al.()Choi, Hong, Park, Chung, and Kim}]{choi_creativeconnect_2024}
{DaEun} Choi, Sumin Hong, Jeongeon Park, John Joon~Young Chung, and Juho Kim.
\newblock \href {https://doi.org/10.1145/3613904.3642794} {{CreativeConnect}: Supporting reference recombination for graphic design ideation with generative {AI}}.
\newblock pages 1--25.
\newblock Conference Name: {CHI} '24: {CHI} Conference on Human Factors in Computing Systems {ISBN}: 9798400703300 Place: Honolulu {HI} {USA} Publisher: {ACM}.

\bibitem[{Cohan et~al.()Cohan, Feldman, Beltagy, Downey, and Weld}]{cohan_specter_2020}
Arman Cohan, Sergey Feldman, Iz~Beltagy, Doug Downey, and Daniel Weld.
\newblock \href {https://doi.org/10.18653/v1/2020.acl-main.207} {{SPECTER}: Document-level representation learning using citation-informed transformers}.
\newblock pages 2270--2282.
\newblock Conference Name: Proceedings of the 58th Annual Meeting of the Association for Computational Linguistics Place: Online Publisher: Association for Computational Linguistics.

\bibitem[{Eger et~al.()Eger, Rücklé, and Gurevych}]{eger_pitfalls_2019}
Steffen Eger, Andreas Rücklé, and Iryna Gurevych.
\newblock \href {https://doi.org/10.18653/v1/W19-4308} {Pitfalls in the evaluation of sentence embeddings}.
\newblock pages 55--60.
\newblock Conference Name: Proceedings of the 4th Workshop on Representation Learning for {NLP} ({RepL}4NLP-2019) Place: Florence, Italy Publisher: Association for Computational Linguistics.

\bibitem[{Gao et~al.()Gao, Chen, Zhao, Liu, Li, Wang, Wang, Guo, and Tang}]{gao_llm4rerank_2025}
Jingtong Gao, Bo~Chen, Xiangyu Zhao, Weiwen Liu, Xiangyang Li, Yichao Wang, Wanyu Wang, Huifeng Guo, and Ruiming Tang.
\newblock \href {https://doi.org/10.1145/3696410.3714922} {{LLM}4rerank: {LLM}-based auto-reranking framework for recommendations}.
\newblock pages 228--239.
\newblock Conference Name: {WWW} '25: The {ACM} Web Conference 2025 {ISBN}: 9798400712746 Place: Sydney {NSW} Australia Publisher: {ACM}.

\bibitem[{Gupta and Pruthi()}]{gupta_all_2025}
Tarun Gupta and Danish Pruthi.
\newblock \href {https://doi.org/10.48550/ARXIV.2502.16487} {All that glitters is not novel: Plagiarism in {AI} generated research}.
\newblock Publisher: {arXiv} Version Number: 2.

\bibitem[{Gómez-Pérez et~al.()Gómez-Pérez, García-Silva, Leone, Albani, Fontaine, Poncet, Summerer, Donati, Roma, and Scaglioni}]{gomez-perez_artificial_2022}
José~Manuel Gómez-Pérez, Andrés García-Silva, Rosemarie Leone, Mirko Albani, Moritz Fontaine, Charles Poncet, Leopold Summerer, Alessandro Donati, Ilaria Roma, and Stefano Scaglioni.
\newblock \href {https://doi.org/10.48550/ARXIV.2210.03640} {Artificial intelligence and natural language processing and understanding in space: A methodological framework and four {ESA} case studies}.
\newblock Publisher: {arXiv} Version Number: 2.

\bibitem[{Joshi et~al.()Joshi, Shahid, Venneti, Vasu, Zheng, Li, Krishnamurthy, and Chan}]{joshi_coprompter_2025}
Ishika Joshi, Simra Shahid, Shreeya~Manasvi Venneti, Manushree Vasu, Yantao Zheng, Yunyao Li, Balaji Krishnamurthy, and Gromit Yeuk-Yin Chan.
\newblock \href {https://doi.org/10.1145/3708359.3712102} {{CoPrompter}: User-centric evaluation of {LLM} instruction alignment for improved prompt engineering}.
\newblock pages 341--365.
\newblock Conference Name: {IUI} '25: 30th International Conference on Intelligent User Interfaces {ISBN}: 9798400713064 Place: Cagliari Italy Publisher: {ACM}.

\bibitem[{Kang et~al.({\natexlab{a}})Kang, Lin, Martelaro, Kittur, Chen, and Hong}]{kang_biospark_2024}
Hyeonsu~B Kang, David Chuan-En Lin, Nikolas Martelaro, Aniket Kittur, Yan-Ying Chen, and Matthew~K. Hong. {\natexlab{a}}.
\newblock \href {https://doi.org/10.1145/3613905.3651035} {{BioSpark}: An end-to-end generative system for biological-analogical inspirations and ideation}.
\newblock pages 1--13.
\newblock Conference Name: {CHI} '24: {CHI} Conference on Human Factors in Computing Systems {ISBN}: 9798400703317 Place: Honolulu {HI} {USA} Publisher: {ACM}.

\bibitem[{Kang et~al.({\natexlab{b}})Kang, Qian, Hope, Shahaf, Chan, and Kittur}]{kang_augmenting_2022}
Hyeonsu~B. Kang, Xin Qian, Tom Hope, Dafna Shahaf, Joel Chan, and Aniket Kittur. {\natexlab{b}}.
\newblock \href {https://doi.org/10.1145/3530013} {Augmenting scientific creativity with an analogical search engine}.
\newblock 29(6):1--36.

\bibitem[{Khattab et~al.()Khattab, Singhvi, Maheshwari, Zhang, Santhanam, Vardhamanan, Haq, Sharma, Joshi, Moazam, Miller, Zaharia, and Potts}]{khattab_dspy_2023}
Omar Khattab, Arnav Singhvi, Paridhi Maheshwari, Zhiyuan Zhang, Keshav Santhanam, Sri Vardhamanan, Saiful Haq, Ashutosh Sharma, Thomas~T. Joshi, Hanna Moazam, Heather Miller, Matei Zaharia, and Christopher Potts.
\newblock \href {https://doi.org/10.48550/ARXIV.2310.03714} {{DSPy}: Compiling declarative language model calls into self-improving pipelines}.
\newblock Publisher: {arXiv} Version Number: 1.

\bibitem[{Kinney et~al.()Kinney, Anastasiades, Authur, Beltagy, Bragg, Buraczynski, Cachola, Candra, Chandrasekhar, Cohan, Crawford, Downey, Dunkelberger, Etzioni, Evans, Feldman, Gorney, Graham, Hu, Huff, King, Kohlmeier, Kuehl, Langan, Lin, Liu, Lo, Lochner, {MacMillan}, Murray, Newell, Rao, Rohatgi, Sayre, Shen, Singh, Soldaini, Subramanian, Tanaka, Wade, Wagner, Wang, Wilhelm, Wu, Yang, Zamarron, Van~Zuylen, and Weld}]{kinney_semantic_2023}
Rodney Kinney, Chloe Anastasiades, Russell Authur, Iz~Beltagy, Jonathan Bragg, Alexandra Buraczynski, Isabel Cachola, Stefan Candra, Yoganand Chandrasekhar, Arman Cohan, Miles Crawford, Doug Downey, Jason Dunkelberger, Oren Etzioni, Rob Evans, Sergey Feldman, Joseph Gorney, David Graham, Fangzhou Hu, and 29 others.
\newblock \href {https://doi.org/10.48550/ARXIV.2301.10140} {The semantic scholar open data platform}.
\newblock Publisher: {arXiv} Version Number: 2.

\bibitem[{Li et~al.()Li, Xu, Guo, Zhao, Li, Yuan, Zhang, Jiang, Xin, Dang, Zhao, Rong, Feng, and Bing}]{li_chain_2024}
Long Li, Weiwen Xu, Jiayan Guo, Ruochen Zhao, Xingxuan Li, Yuqian Yuan, Boqiang Zhang, Yuming Jiang, Yifei Xin, Ronghao Dang, Deli Zhao, Yu~Rong, Tian Feng, and Lidong Bing.
\newblock \href {https://doi.org/10.48550/arXiv.2410.13185} {Chain of ideas: Revolutionizing research via novel idea development with {LLM} agents}.
\newblock \emph{Preprint}, arxiv:2410.13185 [cs].

\bibitem[{Loya et~al.()Loya, Sinha, and Futrell}]{loya_exploring_2023}
Manikanta Loya, Divya Sinha, and Richard Futrell.
\newblock \href {https://doi.org/10.18653/v1/2023.findings-emnlp.241} {Exploring the sensitivity of {LLMs}’ decision-making capabilities: Insights from prompt variations and hyperparameters}.
\newblock In \emph{Findings of the Association for Computational Linguistics: {EMNLP} 2023}, pages 3711--3716. Association for Computational Linguistics.

\bibitem[{Lu et~al.()Lu, Lu, Lange, Foerster, Clune, and Ha}]{lu_ai_2024}
Chris Lu, Cong Lu, R.~T. Lange, J.~Foerster, Jeff Clune, and David Ha.
\newblock \href {https://www.semanticscholar.org/paper/The-AI-Scientist%3A-Towards-Fully-Automated-Discovery-Lu-Lu/33161a5a9b5dcb635b5a97475e6a6209a69ada7d} {The {AI} scientist: Towards fully automated open-ended scientific discovery}.

\bibitem[{Mahajan et~al.()Mahajan, Freestone, Aakur, and Karmaker}]{mahajan_revisiting_2025}
Yash Mahajan, Matthew Freestone, Sathyanarayanan Aakur, and Santu Karmaker.
\newblock \href {https://doi.org/10.48550/ARXIV.2502.19607} {Revisiting word embeddings in the {LLM} era}.
\newblock Publisher: {arXiv} Version Number: 2.

\bibitem[{Meng et~al.()Meng, Arabzadeh, Askari, Aliannejadi, and De~Rijke}]{meng_ranked_2024}
Chuan Meng, Negar Arabzadeh, Arian Askari, Mohammad Aliannejadi, and Maarten De~Rijke.
\newblock \href {https://doi.org/10.1145/3626772.3657864} {Ranked list truncation for large language model-based re-ranking}.
\newblock pages 141--151.
\newblock Conference Name: {SIGIR} 2024: The 47th International {ACM} {SIGIR} Conference on Research and Development in Information Retrieval {ISBN}: 9798400704314 Place: Washington {DC} {USA} Publisher: {ACM}.

\bibitem[{Mysore et~al.({\natexlab{a}})Mysore, Cohan, and Hope}]{mysore_multi-vector_2022}
Sheshera Mysore, Arman Cohan, and Tom Hope. {\natexlab{a}}.
\newblock \href {https://doi.org/10.18653/v1/2022.naacl-main.331} {Multi-vector models with textual guidance for fine-grained scientific document similarity}.
\newblock pages 4453--4470.
\newblock Conference Name: Proceedings of the 2022 Conference of the North American Chapter of the Association for Computational Linguistics: Human Language Technologies Place: Seattle, United States Publisher: Association for Computational Linguistics.

\bibitem[{Mysore et~al.({\natexlab{b}})Mysore, O'Gorman, {McCallum}, and Zamani}]{mysore_csfcube_2021}
Sheshera Mysore, Timothy~J. O'Gorman, A.~{McCallum}, and Hamed Zamani. {\natexlab{b}}.
\newblock \href {https://www.semanticscholar.org/paper/CSFCube-A-Test-Collection-of-Computer-Science-for-Mysore-O'Gorman/6a4deeb40aed8a4d56c8d9401c94b6c7a769e8c3} {{CSFCube} - a test collection of computer science research articles for faceted query by example}.

\bibitem[{Naik et~al.()Naik, Parasa, Feldman, Wang, and Hope}]{naik_literature-augmented_2022}
Aakanksha Naik, Sravanthi Parasa, Sergey Feldman, Lucy Wang, and Tom Hope.
\newblock \href {https://doi.org/10.18653/v1/2022.findings-naacl.33} {Literature-augmented clinical outcome prediction}.
\newblock pages 438--453.
\newblock Conference Name: Findings of the Association for Computational Linguistics: {NAACL} 2022 Place: Seattle, United States Publisher: Association for Computational Linguistics.

\bibitem[{Nouriinanloo and Lamothe()}]{nouriinanloo_re-ranking_2024}
Baharan Nouriinanloo and Maxime Lamothe.
\newblock \href {https://doi.org/10.48550/ARXIV.2406.18740} {Re-ranking step by step: Investigating pre-filtering for re-ranking with large language models}.
\newblock Publisher: {arXiv} Version Number: 1.

\bibitem[{Picard et~al.()Picard, Edwards, Doris, Man, Giannone, Alam, and Ahmed}]{picard_concept_2023}
Cyril Picard, Kristen~M. Edwards, Anna~C. Doris, Brandon Man, Giorgio Giannone, Md~Ferdous Alam, and Faez Ahmed.
\newblock \href {https://doi.org/10.48550/ARXIV.2311.12668} {From concept to manufacturing: Evaluating vision-language models for engineering design}.
\newblock Publisher: {arXiv} Version Number: 3.

\bibitem[{Portenoy et~al.()Portenoy, Radensky, West, Horvitz, Weld, and Hope}]{portenoy_bursting_2022}
Jason Portenoy, Marissa Radensky, Jevin~D West, Eric Horvitz, Daniel~S Weld, and Tom Hope.
\newblock \href {https://doi.org/10.1145/3491102.3501905} {Bursting scientific filter bubbles: Boosting innovation via novel author discovery}.
\newblock pages 1--13.
\newblock Conference Name: {CHI} '22: {CHI} Conference on Human Factors in Computing Systems {ISBN}: 9781450391573 Place: New Orleans {LA} {USA} Publisher: {ACM}.

\bibitem[{Radensky et~al.()Radensky, Shahid, Fok, Siangliulue, Hope, and Weld}]{radensky_scideator_2024}
Marissa Radensky, Simra Shahid, Raymond Fok, Pao Siangliulue, Tom Hope, and Daniel~S. Weld.
\newblock \href {https://doi.org/10.48550/ARXIV.2409.14634} {Scideator: Human-{LLM} scientific idea generation grounded in research-paper facet recombination}.
\newblock Publisher: {arXiv} Version Number: 4.

\bibitem[{Reimers and Gurevych()}]{reimers_sentence-bert_2019}
Nils Reimers and Iryna Gurevych.
\newblock \href {https://doi.org/10.18653/v1/D19-1410} {Sentence-{BERT}: Sentence embeddings using siamese {BERT}-networks}.
\newblock In \emph{Proceedings of the 2019 Conference on Empirical Methods in Natural Language Processing and the 9th International Joint Conference on Natural Language Processing ({EMNLP}-{IJCNLP})}, pages 3980--3990. Association for Computational Linguistics.

\bibitem[{Sarica et~al.()Sarica, Luo, and Wood}]{sarica_technet_2020}
Serhad Sarica, Jianxi Luo, and Kristin~L. Wood.
\newblock \href {https://doi.org/10.1016/j.eswa.2019.112995} {{TechNet}: Technology semantic network based on patent data}.
\newblock 142:112995.

\bibitem[{Sclar et~al.()Sclar, Choi, Tsvetkov, and Suhr}]{sclar_quantifying_2023}
Melanie Sclar, Yejin Choi, Yulia Tsvetkov, and Alane Suhr.
\newblock \href {https://doi.org/10.48550/ARXIV.2310.11324} {Quantifying language models' sensitivity to spurious features in prompt design or: How i learned to start worrying about prompt formatting}.
\newblock Publisher: {arXiv} Version Number: 2.

\bibitem[{Si et~al.()Si, Yang, and Hashimoto}]{si_can_2024}
Chenglei Si, Diyi Yang, and Tatsunori Hashimoto.
\newblock \href {https://www.semanticscholar.org/paper/Can-LLMs-Generate-Novel-Research-Ideas-A-Human-with-Si-Yang/110f5dc6d5bfe67138d64c261d6851c727021d1f} {Can {LLMs} generate novel research ideas? a large-scale human study with 100+ {NLP} researchers}.

\bibitem[{Singh et~al.()Singh, Chang, Anastasiades, Haddad, Naik, Tanaka, Zamarron, Nguyen, Hwang, Dunkleberger, Latzke, Rao, Lochner, Evans, Kinney, Weld, Downey, and Feldman}]{singh_ai2_2025}
Amanpreet Singh, Joseph~Chee Chang, Chloe Anastasiades, Dany Haddad, Aakanksha Naik, Amber Tanaka, Angele Zamarron, Cecile Nguyen, Jena~D. Hwang, Jason Dunkleberger, Matt Latzke, Smita Rao, Jaron Lochner, Rob Evans, Rodney Kinney, Daniel~S. Weld, Doug Downey, and Sergey Feldman.
\newblock \href {https://doi.org/10.48550/arXiv.2504.10861} {Ai2 scholar {QA}: Organized literature synthesis with attribution}.
\newblock \emph{Preprint}, arxiv:2504.10861 [cs].

\bibitem[{Srinivasan and Chan()}]{srinivasan_improving_2024}
Arvind Srinivasan and Joel Chan.
\newblock \href {https://doi.org/10.1145/3635636.3656207} {Improving selection of analogical inspirations through chunking and recombination}.
\newblock pages 374--397.
\newblock Conference Name: C\&C '24: Creativity and Cognition {ISBN}: 9798400704857 Place: Chicago {IL} {USA} Publisher: {ACM}.

\bibitem[{Stevenson and Merlo()}]{stevenson_beyond_2022}
Suzanne Stevenson and Paola Merlo.
\newblock \href {https://doi.org/10.3389/frai.2022.796741} {Beyond the benchmarks: Toward human-like lexical representations}.
\newblock 5:796741.

\bibitem[{Su et~al.()Su, Chen, Tang, Yin, Zheng, Li, Qi, Wu, Li, Ouyang, Torr, Zhou, and Dong}]{su_many_2024}
Haoyang Su, Renqi Chen, Shixiang Tang, Zhenfei Yin, Xinzhe Zheng, Jinzhe Li, Biqing Qi, Qi~Wu, Hui Li, Wanli Ouyang, Philip Torr, Bowen Zhou, and Nanqing Dong.
\newblock \href {https://www.semanticscholar.org/paper/Many-Heads-Are-Better-Than-One%3A-Improved-Scientific-Su-Chen/2e2b5d3589f31cdc5ca0bcd918b22794eb4fe5e4} {Many heads are better than one: Improved scientific idea generation by a {LLM}-based multi-agent system}.

\bibitem[{Suh et~al.()Suh, Chen, Min, Li, and Xia}]{suh_luminate_2024}
Sangho Suh, Meng Chen, Bryan Min, Toby Jia-Jun Li, and Haijun Xia.
\newblock \href {https://doi.org/10.1145/3613904.3642400} {Luminate: Structured generation and exploration of design space with large language models for human-{AI} co-creation}.
\newblock pages 1--26.
\newblock Conference Name: {CHI} '24: {CHI} Conference on Human Factors in Computing Systems {ISBN}: 9798400703300 Place: Honolulu {HI} {USA} Publisher: {ACM}.

\bibitem[{Sun et~al.()Sun, Yan, Ma, Wang, Ren, Chen, Yin, and Ren}]{sun_is_2023}
Weiwei Sun, Lingyong Yan, Xinyu Ma, Shuaiqiang Wang, Pengjie Ren, Zhumin Chen, Dawei Yin, and Zhaochun Ren.
\newblock \href {https://doi.org/10.48550/ARXIV.2304.09542} {Is {ChatGPT} good at search? investigating large language models as re-ranking agents}.
\newblock Publisher: {arXiv} Version Number: 3.

\bibitem[{Wang et~al.({\natexlab{a}})Wang, Wang, Wang, Naidu, Bergen, and Paturi}]{wang_doris-mae_2023}
Jianyou Wang, Kaicheng Wang, Xiaoyue Wang, Prudhviraj Naidu, Leon Bergen, and Ramamohan Paturi. {\natexlab{a}}.
\newblock \href {https://doi.org/10.48550/ARXIV.2310.04678} {{DORIS}-{MAE}: Scientific document retrieval using multi-level aspect-based queries}.
\newblock Publisher: {arXiv} Version Number: 3.

\bibitem[{Wang et~al.({\natexlab{b}})Wang, Dong, and Ma}]{wang_towards_2019}
Kai Wang, Boxiang Dong, and Junjie Ma. {\natexlab{b}}.
\newblock \href {https://doi.org/10.24251/HICSS.2019.111} {Towards computational assessment of idea novelty}.

\bibitem[{Wang et~al.({\natexlab{c}})Wang, Gu, Zhang, Luo, Dai, Shen, Xie, Lin, He, and Ye}]{wang_scipip_2024}
Wenxiao Wang, Lihui Gu, Liye Zhang, Yunxiang Luo, Yi~Dai, Chen Shen, Liang Xie, Binbin Lin, Xiaofei He, and Jieping Ye. {\natexlab{c}}.
\newblock \href {https://doi.org/10.48550/ARXIV.2410.23166} {{SciPIP}: An {LLM}-based scientific paper idea proposer}.
\newblock Publisher: {arXiv} Version Number: 2.

\bibitem[{Xu et~al.()Xu, Bai, Bian, Gao, Wang, Liu, and Liu}]{xu_rc-net_2014}
Chang Xu, Yalong Bai, Jiang Bian, Bin Gao, Gang Wang, Xiaoguang Liu, and Tie-Yan Liu.
\newblock \href {https://doi.org/10.1145/2661829.2662038} {{RC}-{NET}: A general framework for incorporating knowledge into word representations}.
\newblock pages 1219--1228.
\newblock Conference Name: {CIKM} '14: 2014 {ACM} Conference on Information and Knowledge Management {ISBN}: 9781450325981 Place: Shanghai China Publisher: {ACM}.

\bibitem[{Yuksekgonul et~al.()Yuksekgonul, Bianchi, Boen, Liu, Huang, Guestrin, and Zou}]{yuksekgonul_textgrad_2024}
Mert Yuksekgonul, Federico Bianchi, Joseph Boen, Sheng Liu, Zhi Huang, Carlos Guestrin, and James Zou.
\newblock \href {https://doi.org/10.48550/ARXIV.2406.07496} {{TextGrad}: Automatic "differentiation" via text}.
\newblock Publisher: {arXiv} Version Number: 1.

\bibitem[{Zhou et~al.()Zhou, Shen, Geng, Tao, Xu, Long, Jiao, and Jiang}]{zhou_towards_2022}
Yucheng Zhou, Tao Shen, Xiubo Geng, Chongyang Tao, Can Xu, Guodong Long, Binxing Jiao, and Daxin Jiang.
\newblock \href {https://doi.org/10.48550/ARXIV.2206.08063} {Towards robust ranker for text retrieval}.
\newblock Publisher: {arXiv} Version Number: 1.

\end{thebibliography}
